# Non-rigid image registration using spatially region-weighted correlation ratio and GPU-acceleration

Lun Gong, Cheng Zhang, Luwen Duan, Xueying Du, Hanqiu Liu, Xinjian Chen*, Jian Zheng*

*Abstract*—Objective: Non-rigid image registration with high accuracy and efficiency is still a challenging task for medical image analysis. In this work, we present the spatially region-weighted correlation ratio (SRWCR) as a novel similarity measure to improve the registration performance. Methods: SRWCR is rigorously deduced from a three-dimension joint probability density function combining the intensity channels with an extra spatial information channel. SRWCR estimates the optimal functional dependence between the intensities for each spatial bin, in which the spatial distribution modeled by a cubic B-spline function is used to differentiate the contribution of voxels. We also analytically derive the gradient of SRWCR with respect to the transformation parameters and optimize it using a quasi-Newton approach. Furthermore, we propose a GPU-based parallel mechanism to accelerate the computation of SRWCR and its derivatives. Results: The experiments on synthetic images, public 4-D thoracic computed tomography (CT) dataset, retinal optical coherence tomography (OCT) data, and clinical CT and positron emission tomography (PET) images confirm that SRWCR significantly outperforms some state-of-the-art techniques such as spatially encoded mutual information and Robust PaTch-based cOrrelation Ration. Conclusion: This study demonstrates the advantages of SRWCR in tackling the practical difficulties due to distinct intensity changes, serious speckle noise, or different imaging modalities. Significance: The proposed registration framework might be more reliable to correct the non-rigid deformations and more potential for clinical applications.

The paper is submitted for review on April 3, 2018. This work was supported in part by the National Program on Key Research and Development Project under Grants 2016YFC0104500, 2016YFC0104505, 2016YFC0103500, 2016YFC0103502, in part by NSFC under Grants 61201117, 61701492, in part by the NSFJ under Grants BK20151232, BK20170392, and in part by the Youth Innovation Promotion Association CAS under Grant 2014281.

L. Gong, C. Zhang, L. Duan, X. Du and *J. Zheng are with Medical Imaging Department, Suzhou Institute of Biomedical Engineering and Technology, Chinese Academy of Sciences, Suzhou, 215163 China (e-mail: gonglun@hotmail.com; zhangc@sibet.ac.cn; dlwj0222@mail.ustc.edu.cn ; duxy_1993@163.com; e-mail: zhengj@sibet.ac.cn).
H. Liu is with Department of Radiology, Huashan hospital, Fudan University, Shanghai, 200000 China (e-mail: drhancher@163.com).
*X. Chen is with the School of Electronics and Information Engineering, Soochow University, Suzhou, 215163 China(e-mail: xjchen@suda.edu.cn).
Corresponding authors: *J. Zheng , *X. Chen.

## I. INTRODUCTION

Non-rigid image registration plays a more and more important role in a variety of applications, such as radiation assessment [1], disease surveillance [2], atlas-based segmentation [3], image-guided surgery [4] and so on. All these applications are based on plausibly correcting the spatial distortion between the corresponding anatomical tissues from different images. Over the nearly two decades, lots of non-rigid registration approaches have been proposed to estimate the dense deformation fields [5], [6].

Currently, the most widely-used method estimates the alignment correspondences by optimizing an intensity-based metric. Mutual information (MI) [7], [8] is a popular metric and has been successfully applied to both mono-modal and multi-modal registration [9]. It is derived from information theory and quantifies the amount of statistical information that one image depends on the other. However, several recent studies have confirmed that the optimal alignment might be not corresponding to the hypothesis of intensity bin correspondence held by MI-based registration, especially for aligning images with intensity distortion or different modalities [10]-[12]. Embedding spatial information into the calculation of MI is an effective approach to improve the registration accuracy [13]-[19].

The most direct way was to estimate a high-dimensional MI by considering the neighbors of each voxel [13], [14]. This method required plenty of samples to ensure the accuracy of the joint entropy. Some studies incorporated the spatial information by combining MI with geometric features, such as image gradient [15] and 3D Harris operator [16]. However, for multi-modal registration, particularly when the image pairs have absolutely different representations for the same tissues, it is quite challenging to detect enough features in both two images. Recent approaches proposed to weight the MI metric using local structural information [11] or contextual similarities [12]. However, the computational complexity increased significantly due to the calculation of self-similarity or the detection of similar structures.

Another strategy was to extend the 2-D intensity joint histogram with an extra spatial channel representing the location of the intensity pairs. For each spatial bin, this three-channel strategy calculated a local MI value with a given spatial distribution. Studholme *et al.* [17] firstly introduced this strategy to encode the location information and proposed regional mutual information (RMI) as a novel similarity measure. RMI manually partitioned the image into a set of overlapping squares and mapped the spatial distribution to a boxcar function. The experiments on brain MR images showed that RMI was more robust to local intensity changes. Instead of subdividing regions over the whole image domain, localized mutual information (LMI) [18] provided a random partition mechanism where subregions were the neighborhoods of lots of stochastic points. However, both RMI and LMI assumed that the weights of voxels within each subregion were equal, regardless of their different coordinates. Conditional mutual information (CMI) [19] addressed this problem by fitting the spatial distribution to a tensor-product B-spline function. For each subregion, the weight of each voxel was associated with the distance to the center. The experiments on simulated, mono-modal and multi-modal data demonstrated that CMI obviously improved the registration accuracy compared with MI and RMI. Following the weighting scheme of CMI, spatially encoded mutual information (SEMI) [10] exploited the Gaussian function for spatial distribution and divided the image into a set of spheres. To reduce the registration time, a local ascent optimization framework was presented to minimize SEMI. Essentially, SEMI-based registration improved the efficiency at the cost of precision. It estimated the approximate derivatives using the voxels with larger weights, which greatly deteriorated the global convergence.

Although these three-channel MI metrics have achieved satisfying performance for many applications, especially for the cases with spatial intensity changes, there are still some obstacles. One is the conflict between the reliable MI estimations and the local performance as for the regional size. In general, MI requires enough samples to ensure the statistical power, which means that the regional size should be as large as possible. However, the literature [10] concludes that the smaller the size of each subregion is, the better locality of the three-channel metric will be. To maintain the balance between the two advantages, SEMI-based registration employed a hierarchical weighting scheme which halved the regional size at each subsequent level. But this scheme did not solve the problem fundamentally. Additionally, due to the huge computational burden of MI [20], the computing time to estimate a series of local MI values is not practical.

Correlation ratio (CR), first introduced by Roche *et al.* [21], [22], is an alternative intensity-based metric for image registration. With the hypothesis of a more restrictive functional mapping between the intensities, it is demonstrated to be with less computational complexity and more reliable to small sample sets [21], [22]. Recently, Rivaz *et al.* [23] incorporated the three-channel strategy into CR and presented Robust PaTch-based cOrrelation Ration (RaPTOR) algorithm, which estimated local CR values over small patches and accumulated these values to constitute a global cost function. Similar to LMI, RaPTOR randomly selected patches as spatial bins and fitted the spatial distribution to a boxcar function. Compared with LMI, it was more reliable in handling large intensity distortion due to the better locality performance.

In this paper, we incorporate the spatial information of voxels into the functional dependence between the intensities and present the spatially region-weighted correlation ratio (SRWCR) as a novel similarity measure. The work of RaPTOR is acted as a starting point. RaPTOR maps the spatial distribution to a boxcar function, which is not accurate to express the location of the joint intensity pairs (see Section II-C). Additionally, RaPTOR employs a frequency statistics to simplify the derivation, which significantly reduces the stability for different datasets. In contrast, our proposed SRWCR is rigorously deduced from a 3-D joint probability density function (PDF). For each spatial bin, the contribution of different voxels is distinguished by a cubic B-spline function. Consequently, SRWCR is more effective in estimating the optimal mapping relationship between the intensities within each subregion. SRWCR not only inherits the robustness of RaPTOR to intensity distortion, but also becomes less sensitive to speckle noise and more suitable to align images with different modalities.

In our previous work [24], we have obtained some preliminary results of SRWCR-based registration on a small retinal OCT dataset. Here, we extend this work as follows: 1) we investigate the three-channel strategy using different spatial distributions and provide a more principled description of SRWCR and its derivatives; 2) we combine the CUDA-programming with the three-channel strategy and propose a GPU-based highly efficient framework to speed up the calculation; 3) we perform more comprehensive assessments on four different datasets, including 3-D synthetic images distorted with initial bias fields, public 4-D thoracic CT scans with distinct intensity changes, 3-D retinal OCT images disturbed by strong speckle noise, and 3-D lung CT/PET scans.

The rest of the paper is organized as follows. In the next section, we will give a detailed description of our method. The registration performance is then validated on several datasets in Section III, followed by a profound discussion presented in Section IV. And the final summary is listed in Section V.

## II. METHODS

Let $\Omega = \{\mathbf{x}=(x,y,z) | 0 \leq x < N_x, 0 \leq y < N_y, 0 \leq z < N_z\} \subset \mathbb{R}^3$ denote the whole image domain, and then aligning a moving image $M$ to a fixed image $F$ can be formulated as an optimization problem which minimizes a cost function defined by

$$C = D(F(\mathbf{x}), M(T(\mathbf{x};\Phi))) + w_p \times C_p(T(\mathbf{x};\Phi)) \quad (1)$$

where $D$ is a similarity measure, and $C_p$ is the constraint of the deformation fields that avoids implausible movements. $w_p$ is an experiential penalty weight and $T(\mathbf{x};\Phi)$ represents the transformation parameterized by $\Phi$. We choose a free-form deformation (FFD) modeled by the location of cubic B-spline nodes to simulate the transformation. The spacings of the control nodes along the three directions are respectively set to $\delta_x, \delta_y, \delta_z$. Therefore, $\Phi = (\phi_0, \phi_1, \cdots \phi_n)$ is a set of coordinates of all control nodes. In this work, we mainly focus on the construction of the accelerated-SRWCR (A-SRWCR). To simplify the calculation, the intensities of the fixed and moving images have been normalized between zero and the maximal intensity bin $L_\varepsilon$. The flowchart of our algorithm, which adopts a GPU-based acceleration scheme, is shown in Fig. 1.

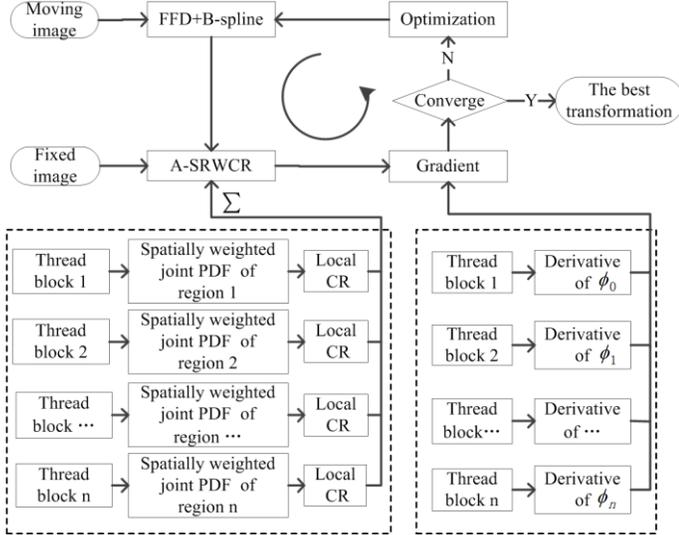

Fig. 1. Flowchart of our A-SRWCR based registration framework. The two dashed boxes indicate the parallel mechanism of the novel similarity measure and its derivatives, respectively.

### A. Correlation Ratio (CR)

CR is first introduced to the rigid alignment of multi-modal images by Roche et al. [21], [22]. For two images $A$ and $B$, CR assumes that all intensities of $B$ can be estimated by $A$ using an unknown function, and it measures the dispersion of the mapping relationship from this function.

Supposing that $\sigma^2$ is the variance of the estimated image $B$ and $p(a)$ denotes the marginal PDF of the model image $A$, and then CR takes the following form

$$CR(A,B) = 1 - \frac{1}{\sigma^2} \sum_{a=0}^{L_\varepsilon} \sigma^2(a)\, p(a) \qquad (2)$$

where $a \in \{0, 1, 2, \cdots, L_\varepsilon\}$ represents the discrete intensity bins associated to $A$, and $\sigma^2(a)$ is the conditional variance of $B$ given a specific intensity bin $a$. CR varies between 0 and 1: the higher the value is, the more perfect the functional dependence is. Unlike MI, CR is asymmetric noted as $CR(A,B) \neq CR(B,A)$. Therefore, it is important to choose the proper estimated image $B$ from $F$ and $M$.

### B. Three-Dimension Joint Density Function

The major drawback of CR is that it establishes a purely quantitative functional dependence between the intensities, but ignores the spatial information shared across the image. Following the three-channel strategy, we divide the whole image domain into a series of subregions labeled by $r \in R = \{0, 1, 2, \cdots, R_n\}$, and assume that the possibility of voxel $\mathbf{x}$ appearing in region $r$ is corresponding to a spatial distribution defined by $w(r,\mathbf{x})$. Hence, a 3-D PDF can be defined using a weighted intensity statistics scheme

$$p(a,b,r) = \frac{1}{Z} \sum_{\mathbf{x} \in \Omega} w(r,\mathbf{x}) h(a - A(\mathbf{x})) h(b - B(\mathbf{x})) \qquad (3)$$

where $a \in \{0, 1, 2, \cdots, L_\varepsilon\}$ and $b \in \{0, 1, 2, \cdots, L_\varepsilon\}$, and $r$ can be viewed as the spatial bin corresponding to the label of subregion. $Z$ is a normalization factor. In essence, $p(a,b,r)$ represents the probability of a pair of intensity bins $a$ and $b$ co-occurring in region $r$. According to the characteristics of the conditional probability, it can be re-written as the product of the probability of intensity pairs $(a,b)$ within a given subregion $r$, called $p_r(a,b)$, and the probability $p(r)$ that region $r$ occurs in the whole image domain, so that

$$\begin{aligned} p(r) &= \sum_{a=0}^{L_\varepsilon} \sum_{b=0}^{L_\varepsilon} p(a,b,r) \\ p_r(a,b) &= p(a,b,r)/p(r) \\ p_r(a) &= \sum_{b=0}^{L_\varepsilon} p_r(a,b) \\ p_r(b) &= \sum_{a=0}^{L_\varepsilon} p_r(a,b) \end{aligned} \qquad (4)$$

In this work, a second-order polynomial function designed by Xu et al. [25] which is smoother and more differentiable is used to estimate the joint histogram

$$h(t) = \begin{cases} -1.8|t|^2 - 0.1|t| + 1, & 0 \le |t| < 0.5 \\ 1.8|t|^2 - 3.7|t| + 1.9, & 0.5 \le |t| < 1 \\ 0, & \text{otherwise} \end{cases} \qquad (5)$$

### C. Encoding Location Information from Spatial Distribution

According to (3), there are two key points to integrate the spatial information into the functional dependence: the way to partition the subregions and the specific expression of the spatial distribution. Similar to LMI, RaPTOR randomly selects points as the centers of subregions and maps the spatial distribution to a boxcar function

$$w(r,\mathbf{x}) = \begin{cases} 1, & \mathbf{x} \in \Omega_r \\ 0, & \mathbf{x} \notin \Omega_r \end{cases} \qquad (6)$$

where $\Omega_r$ is the user-defined subregion corresponding to spatial bin $r$. Fig. 2(a) illustrates the spatial distribution with a given spatial bin in a 2-D image domain. RaPTOR assumes that the probabilities of voxels $\mathbf{x} \in \Omega_r$ are equal, without consideration of corresponding spatial coordinates. It therefore discards the topological relationship between these voxels and may lead to unrealistic deformations.

In this work, the cubic B-spline function which has been introduced into CMI is provided for $w(r,\mathbf{x})$, with the same setting as FFD for control nodes and grid spacings. It holds that the weights of voxels $\mathbf{x} \in \Omega_r$ are monotonically decreasing with respect to their distances to the center of region $r$. By taking the control nodes of FFD as the centers and utilizing the $l,m,q$-th degree B-spline basis function in each dimension, $w(r,\mathbf{x})$ can be given by

$$w(r,\mathbf{x}) = \begin{cases} \beta_l^3(x - \phi_{r,x})\beta_m^3(y - \phi_{r,y})\beta_q^3(z - \phi_{r,z}), & \mathbf{x} \in \Omega_r \\ 0, & \mathbf{x} \notin \Omega_r \end{cases} \qquad (7)$$

where $(\phi_{r,x}, \phi_{r,y}, \phi_{r,z})$ is the coordinate of control point $\phi_r$, and $\beta^3$ represents the B-spline function listed in (8). Due to the limited span characters of the cubic B-spline function, the subregion $\Omega_r$ is restrained to a $4\delta_x \times 4\delta_y \times 4\delta_z$ cuboid centered on $\phi_r$, as shown in Fig. 2(b).

$$\begin{cases} \beta_0^3(t) = (1-t)^3/6 \\ \beta_1^3(t) = (3t^3 - 6t^2 + 4)/6 \\ \beta_2^3(t) = (-3t^3 + 3t^2 + 3t + 1)/6 \\ \beta_3^3(t) = t^3/6 \end{cases}. \quad (8)$$

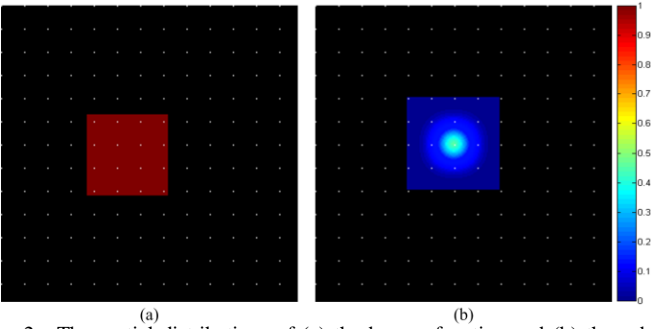

Fig. 2. The spatial distributions of (a) the boxcar function and (b) the cubic B-spline function given an arbitrary spatial bin, respectively.

### D. SRWCR

#### 1) Similarity Measure

SRWCR computes a series of local CR values by estimating the regional statistical properties on each subregion, and adds these local values to quantify how well the two images match. If we treat the labels of subregions as the indexes of spatial bins, SRWCR can be defined by a weighted scheme as follows

$$D(A,B) = SRWCR(A,B,R) = \sum_{r=0}^{Rn} p(r)(1 - CR(A,B|r))$$
$$= \sum_{r=0}^{Rn} p(r)\left(\frac{1}{\sigma_r^2}\sum_{a=0}^{L\varepsilon} \sigma_r^2(a) p_r(a)\right) \quad (9)$$

where $\sigma_r^2, \sigma_r^2(a)$ are the regional variances estimated in a given spatial bin $r$. Referred to [22], they can be derived as follows

$$\sigma_r^2 = \sum_{b=0}^{L\varepsilon} b^2 p_r(b) - \mu_r^2, \quad \mu_r = \sum_{b=0}^{L\varepsilon} b p_r(b),$$
$$\sigma_r^2(a) = \frac{\sum_{b=0}^{L\varepsilon} b^2 p_r(a,b)}{p_r(a)} - \mu_r^2(a), \quad \mu_r(a) = \frac{\sum_{b=0}^{L\varepsilon} b p_r(a,b)}{p_r(a)} \quad (10)$$

where $p_r(a), p_r(b), p_r(a,b)$ are the regional PDFs defined in (4), $\mu_r$ and $\mu_r(a)$ are the expectation and conditional expectation of $B$ estimated within the corresponding subregion $\Omega_r$, respectively.

Integrating (10) into (9), we have

$$SRWCR(A,B,R) = \sum_{r=0}^{Rn} p(r)\left[\frac{1}{\sigma_r^2}\left(\sum_{b=0}^{L\varepsilon} b^2 p_r(b) - \sum_{a=0}^{L\varepsilon} p_r(a)\mu_r^2(a)\right)\right]. \quad (11)$$

Due to the independency of $\mu_r(a)$ and $B$, (11) can be simplified to a more concise form which contains the information of two intensity channels and a spatial channel

$$SRWCR(A,B,R) = \sum_{r=0}^{Rn} p(r) \sum_{a=0}^{L\varepsilon} \sum_{b=0}^{L\varepsilon} \left[\frac{(b^2 - \mu_r^2(a))}{\sigma_r^2} p_r(a,b)\right]$$
$$= \sum_{r=0}^{Rn} \sum_{a=0}^{L\varepsilon} \sum_{b=0}^{L\varepsilon} \left[\frac{(b^2 - \mu_r^2(a))}{\sigma_r^2} p(a,b,r)\right]. \quad (12)$$

SRWCR is a dissimilarity measure and the registration accuracy immensely relies on choosing whether $F$ or $M$ to act as the estimated image $B$. It varies between 0 and 1: for registered images with functional intensity mappings, its value is close to 0. If we exploit a random partitioning scheme with a boxcar function to determine the spatial distribution, SRWCR will degrade into RaPTOR in virtue of some approximate reductions

$$RaPTOR(A,B,R) = \frac{1}{N_p}\sum_{r=0}^{Rn}\frac{1}{\sigma_r^2}\left(\sum_{\mathbf{x}\in\Omega_r}\frac{B^2(\mathbf{x})}{N_r} - \sum_{a=0}^{L\varepsilon} p_r(a)\mu_r^2(a)\right) \quad (13)$$

with

$$p_r(a) = \frac{1}{N_r}\sum_{\mathbf{x}\in\Omega_r} h(a - A(\mathbf{x})), \quad \mu_r(a) = \frac{\sum_{\mathbf{x}\in\Omega_r} h(a - A(\mathbf{x}))B(\mathbf{x})}{N_r p_r(a)}. \quad (14)$$

where $N_p$ is the number of patches, and $N_r$ is the size of each patch. RaPTOR just counts the frequencies to estimate $p_r(b)$ and $\sigma_r^2$. This approximation greatly weakens the differentiable property. In addition, instead of using a second-order function like (5), RaPTOR employs a linear function to estimate $p_r(a)$ and $\mu_r(a)$. It may cause artificial quantization errors of the joint histogram and deteriorate the registration performance.

#### 2) Accelerated-SRWCR

A major limitation of the three-channel strategy is the large computational burden. However, the calculation process of each spatial bin is actually independent. In this section, we propose A-SRWCR to speed up the estimation by means of the graphics processing unit (GPU).

It seems to be an efficient acceleration scheme that the regional joint PDF $p_r(a,b)$ of each spatial bin is stored in shared memory and each thread block directly computes a local CR value. But this scheme contains complex nested loop operations in the thread block, greatly decreasing the acceleration efficiency. Therefore, we separately calculate $p_r(a,b)$ for each spatial bin using CUDA-programming, and the cumbersome loop operations are performed on the CPU. Table I gives the pseudo-code of our proposed A-SRWCR, in which the parallel acceleration can be found in line 1 and 2.

Our CUDA implementation consists of two kernels. The first kernel transforms the moving image in terms of the location of the control nodes. Another kernel generates series of regional joint PDFs as outputs by maintaining a histogram per thread block. According to (7), since only voxels $\mathbf{x}\in\Omega_r$ have contribution to updating $p_r(a,b)$, we assign threads to these voxels instead of the whole image domain. All regional joint PDFs are stored in global memory and the atomic function is employed to prevent the thread conflicts.

Using the proposed A-SRWCR, we can significantly improve the matching performance quantification between two images. In next section, we analytically compute the derivatives of SRWCR to take the advantage of the gradient-based optimization framework.

TABLE I
THE PSEUDO-CODE OF A-SRWCR.

**Input:** the fixed image $F$, the moving image $M$, and the transformation parameters $\Phi$.
**Output:** the value of SRWCR $E$.
1  transform the moving image $M$ in the GPU;
2  compute $p_r(a,b)$ for each spatial bin using (3) and (4) in the GPU;
3  copy the series of regional joint PDFs from GPU to CPU;
4  $E=0$;
5  **for** each spatial bin $r \in R$
6      compute the marginal density functions $p_r(a)$ and $p_r(b)$ using (4);
7      compute the variance $\sigma_r^2$ of the estimated image $B$ (either $F(\mathbf{x})$ or $M(T(\mathbf{x};\mathbf{u}))$) using (10);
8      **for** $a=0:L_\varepsilon$
9          compute the conditional expectation $\mu_r(a)$ with a given intensity bin $a$ using (10);
10     **end for**
11     $cr=0$;
12     **for** $a=0:L_\varepsilon$
13         **for** $b=0:L_\varepsilon$
14             $cr += \left(b^2 - \mu_r^2(a)\right) p_r(a,b) / \sigma_r^2$;
15         **end for**
16     **end for**
17     $E += p(r) \times cr$;
18 **end for**

### E. Gradient

#### 1) Analytical Derivations

For an independent parameter vector $\Phi$, the gradient of SRWCR can be separately computed as follows

$$\nabla D = \left[ \frac{\partial D}{\partial \phi_0}, \frac{\partial D}{\partial \phi_1}, \dots, \frac{\partial D}{\partial \phi_s}, \frac{\partial D}{\partial \phi_n} \right]. \quad (15)$$

By means of the chain rule, a single component of the gradient can be deduced as follows

$$\frac{\partial D}{\partial \phi_s} = \sum_{\mathbf{x} \in \Omega} \left\{ \frac{\partial D}{\partial M(\mathbf{y})}\bigg|_{\mathbf{y}=T(\mathbf{x};\mathbf{u})} \cdot \frac{\partial M(\mathbf{y})}{\partial \mathbf{y}} \cdot \frac{\partial T(\mathbf{x};\phi)}{\phi_s} \right\} \quad (16)$$

where $\partial D/\partial M(\mathbf{y})$ is the derivative of SRWCR with respect to the intensity of current voxel, $\partial M(\mathbf{y})/\partial \mathbf{y}$ is the gradient of the moving image, and $\partial T/\partial \phi_s$ denotes the Jacobian of the transformation parameter as calculated by (17).

$$\frac{\partial T(\mathbf{x};\phi)}{\partial \phi_s} = \begin{cases} \beta_{i-p_x}^3(\eta)\beta_{j-p_y}^3(\gamma)\beta_{k-p_z}^3(\tau), & \mathbf{x} \in V \\ 0, & \mathbf{x} \notin V \end{cases} \quad (17)$$

where $p_x = \lfloor x/\delta_x \rfloor - 1$, $p_y = \lfloor y/\delta_y \rfloor - 1$, $p_z = \lfloor z/\delta_z \rfloor - 1$, $\eta = x/\delta_x - \lfloor x/\delta_x \rfloor$, $\gamma = y/\delta_y - \lfloor y/\delta_y \rfloor$, $\tau = z/\delta_z - \lfloor z/\delta_z \rfloor$, $\beta^3$ is the cubic B-spline function as defined by (8), and $i, j, k$ are the indexes of current control node along the three directions. $V$ denotes the local region influenced by $\phi_s$.

Considering that either $A$ or $B$ can be acted as the moving image, we severally compute the derivatives of (11) with respect to both $A$ and $B$. Firstly, if we view $B$ as $M(\mathbf{y})$, $\partial D/\partial M(\mathbf{y})$ can be deduced as follows

$$\frac{\partial D}{\partial M(\mathbf{y})} = \sum_{r=0}^{R_n} p(r) \left[ \frac{1}{\sigma_r^2} \left( \sum_{b=0}^{L_\varepsilon} b^2 \frac{\partial p_r(b)}{\partial B(\mathbf{y})} - \sum_{a=0}^{L_\varepsilon} 2 p_r(a) \mu_r(a) \frac{\partial \mu_r(a)}{\partial B(\mathbf{y})} \right) \right. \\ \left. + \frac{-\partial \sigma_r^2/\partial B(\mathbf{y})}{\sigma_r^2} (1 - CR(A,B \mid r)) \right] \quad (18)$$

where $CR(A,B \mid r)$ denotes the local CR value as defined in (9).

Referred to the reductions in Appendix I, (18) can be re-written as follows

$$\frac{\partial D}{\partial M(\mathbf{y})} = \frac{1}{Z} \sum_{r=0}^{R_n} \sum_{a=0}^{L_\varepsilon} \sum_{b=0}^{L_\varepsilon} \left[ \frac{(1-CR(A,B \mid r))(b^2 - 2b\mu_r) + 2b\mu_r(a) - b^2}{\sigma_r^2} \right. \\ \left. \cdot w(r,\mathbf{x}) h(a - A(\mathbf{x})) \frac{\partial h}{\partial \kappa}\bigg|_{\kappa=b-B(\mathbf{y})} \right] \quad (19)$$

where $\partial h/\partial \kappa$ represents the first-order derivative of the parzen-window function.

Due to the asymmetric property of SRWCR, $\partial D/\partial M(\mathbf{y})$ is completely different when $A$ is corresponding to the moving image. Similar to the above, we analytically derive the derivatives of (11) with respect to $A$ as follows

$$\frac{\partial D}{\partial M(\mathbf{y})} = \sum_{r=0}^{R_n} \frac{-p(r)}{\sigma_r^2} \left[ \sum_{a=0}^{L_\varepsilon} \left( 2 p_r(a) \mu_r(a) \frac{\partial \mu_r(a)}{\partial A(\mathbf{y})} + \mu_r^2(a) \frac{\partial p_r(a)}{\partial A(\mathbf{y})} \right) \right]. \quad (20)$$

Referred to the reductions in Appendix II, (20) can be re-written as follows

$$\frac{\partial D}{\partial M(\mathbf{y})} = \frac{1}{Z} \sum_{r=0}^{R_n} \sum_{a=0}^{L_\varepsilon} \sum_{b=0}^{L_\varepsilon} \left[ \frac{2b - \mu_r(a)}{\sigma_r^2} \mu_r(a) w(r,\mathbf{x}) h(b - B(\mathbf{x})) \frac{\partial h}{\partial \kappa}\bigg|_{\kappa=a-A(\mathbf{y})} \right]. \quad (21)$$

#### 2) Parallel Optimization

According to (15), the gradient of SRWCR can be separated into a set of independent components, which is convenient to be accelerated using CUDA-programming. The challenge is how to parallel the calculation of (16) effectively.

Our parallel mechanism contains two kernels. The first kernel implements the calculation of $\partial D/\partial M(\mathbf{y})$ and $\partial M(\mathbf{y})/\partial \mathbf{y}$ with respect to all voxels. Benefiting from the limited span characters of the cubic B-spline function, each voxel $\mathbf{x}$ belongs to the nearest 64 cuboids. Thus, we only focus on the corresponding 64 spatial bins to update $\partial D/\partial M(\mathbf{y})$. Additionally, according to the parzen-window function defined by (5), only two nearby intensity bins of $A(\mathbf{x})$ or $B(\mathbf{x})$ have effects on the computation of (19) or (21). Another kernel performs the gradient of SRWCR by maintaining a single component of (15) per thread block. The outputs of the first kernel are stored in global memory as inputs and reused by all thread blocks.

### F. Implementation Details

In this work, the bending energy of the deformation fields is acted as the constraint term $C_p$, referred to [26] for more details. The backward warping is utilized to transform the images and the splatting algorithm with a Gaussian function is employed to invert the deformation fields. The multi-resolution strategy and the concatenation of three isotropic control grids are used to improve search efficiency.

The experiential penalty factor $w_p$ is set to 0.1 for mono-modal registration and 30 for multi-modal registration. The grid spacing at the finest level and the maximal intensity bin are chosen to be fixed for all experiments ($\delta =[5, 5, 5]$ and $L_\varepsilon = 31$).

The minimization of the cost function is performed with the quasi-Newton limited-memory BFGS (LBFGS) provided in the liblbfgs package (http://www.chokkan.org/software/liblbfgs/). The number of Hessian corrections is set to 5, and the maximal iteration is set to 200, 200, 120 for low, medium and high resolution. The backtracking line search with a Wolfe condition is adopted. The optimizer is stopped when the metric value is stable within the last 20 steps or when the optimization reaches the maximal iterations.

All experiments are accomplished on a PC equipped with 8 GB RAM, Intel Core i7 3.4 GHz CPU and an NVIDIA Geforce GTX 1060 graphics card. An executable tool of the presented algorithm is available online at https://github.com/Gonglun/Registration.

### III. Results

To evaluate the performance of A-SRWCR for non-rigid image registration, a number of tests are performed on synthetic images, public 4-D thoracic CT dataset, retinal OCT images and clinical lung CT/PET scans. All experiments are carried out following the principles of the Declaration of Helsinki, and approved by the volunteers and patients for publication.

For each dataset, the proposed method is compared with the classical MI metric and two state-of-the-art three-channel metrics: SEMI and RaPTOR. Among them, the popular registration package Elastix [27] based on MI and the existing executable tool of SEMI can be severally downloaded from the homepages (http://elastix.isi.uu.nl/) and (http://www.sdspeople.fudan.edu.cn/zhuangxiahai/0/zxhproj/), and RaPTOR is accomplished following the idea of [23] using CUDA-programming. For clinical datasets without anatomical landmarks, the Hausdorff distance (HD) which quantifies the maximum distance between two outlines and the maximum-likelihood Hausdorff distance (MHD) [28] which quantifies the mean distance of all points are utilized to evaluate the accuracy of the alignment.

#### A. Registration of synthetic images

We evaluate the robustness of the four metrics to a bias field on ten synthetic image pairs. In this experiment, a 3-D binary black and white grid image with dimension $128\times128\times128$ is aligned with a warped version of itself. For each image pair, the warped image is distorted with a B-spline transformation field, in which the control nodes $\Phi$ are initialized by a uniform distribution with the maximum amplitude of 15 voxels. The example slices illustrating the data are shown in Fig. 3. To quantitatively assess the registration accuracy, we compare the registered displacement of each voxel with the initial by computing the root mean square error (RMSE) over the whole image domain.

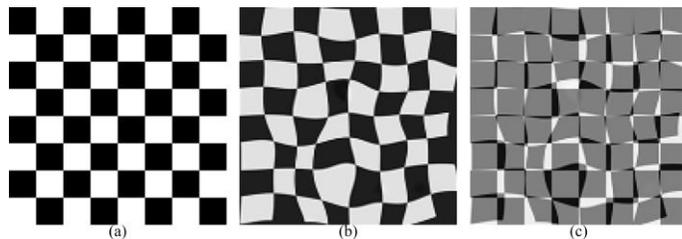

Fig. 3. Synthetic slices, distorted with known ground truth. (a) is the slice from the original image; (b) is the corresponding slice from the warped image; (c) is the difference between (a) and (b).

Due to the asymmetric nature of SRWCR, the moving image $M$ can be set to either $A$ or $B$. Additionally, both the original image $O$ and the warped image $W$ can serve as $M$. Consequently, there are four different combinations. We first investigate the differences among the four combinations. As shown in Table II, the registration performance greatly depends on the combination, and the average RMSE ranges from $0.78\pm0.07$ voxels to $1.89\pm0.08$ voxels. Therefore, in the following experiments, we randomly select a case from each dataset, and compare the accuracy of the four combinations to determine the optimal one.

TABLE II
QUANTITATIVE ANALYSIS OF DIFFERENT COMBINATIONS. M IS THE MOVING IMAGE. A AND B ARE THE MODEL AND ESTIMATED IMAGES. O AND W ARE THE ORIGINAL AND WARPED IMAGES.

| Case | RMSE [voxel] for each combination | | | | |
|---|---|---|---|---|---|
| | Initial | M as B | | M as A | |
| | | W as M | O as M | W as M | O as M |
| #1 | 4.46 | 1.91 | 1.09 | 1.74 | **0.82** |
| #2 | 4.50 | 1.90 | 0.98 | 1.68 | **0.71** |
| #3 | 4.57 | 1.93 | 1.03 | 1.82 | **0.85** |
| #4 | 4.37 | 1.81 | 0.91 | 1.58 | **0.72** |
| #5 | 4.44 | 1.78 | 1.03 | 1.68 | **0.79** |
| #6 | 4.55 | 1.85 | 0.96 | 1.69 | **0.79** |
| #7 | 4.59 | 2.02 | 1.03 | 1.77 | **0.76** |
| #8 | 4.23 | 1.85 | 1.05 | 1.66 | **0.72** |
| #9 | 4.38 | 1.83 | 0.99 | 1.68 | **0.91** |
| #10 | 4.34 | 1.99 | 0.97 | 1.69 | **0.71** |
| Ave | 4.44 | 1.89 | 1.00 | 1.70 | **0.78** |
| Std | 0.11 | 0.08 | 0.05 | 0.07 | **0.07** |

We also compare the registration accuracy of A-SRWCR with the other three metrics. As shown in Table III, our proposed metric achieves the lowest RMSE for each case, and outperforms the second best results ($0.95\pm0.12$ voxels) by 18%. This demonstrates the effectiveness of A-SRWCR in recovering the realistic deformations. Additionally, we also find that MI provides the worst result for each case, which verifies the robustness of encoding spatial information to a bias field.

TABLE III
QUANTITATIVE ANALYSIS OF SYNTHETIC DATASET

| Case | RMSE [voxel] for each method | | | | |
|---|---|---|---|---|---|
| | Initial | MI | SEMI | RaPTOR | A-SRWCR |
| #1 | 4.46 | 1.32 | 1.14 | 1.00 | **0.82** |
| #2 | 4.50 | 1.13 | 1.06 | 0.97 | **0.71** |
| #3 | 4.57 | 1.37 | 1.19 | 0.96 | **0.85** |
| #4 | 4.37 | 1.18 | 1.08 | 0.97 | **0.72** |
| #5 | 4.44 | 1.36 | 1.20 | 1.05 | **0.79** |
| #6 | 4.55 | 1.15 | 1.01 | 0.84 | **0.79** |
| #7 | 4.59 | 1.20 | 0.98 | 0.84 | **0.76** |
| #8 | 4.23 | 1.18 | 1.00 | 0.78 | **0.72** |
| #9 | 4.38 | 1.38 | 1.21 | 1.19 | **0.91** |
| #10 | 4.34 | 1.39 | 1.17 | 0.87 | **0.71** |
| Ave | 4.44 | 1.27 | 1.10 | 0.95 | **0.78** |
| Std | 0.11 | 0.11 | 0.09 | 0.12 | **0.07** |

### B. Registration of extreme inhale and exhale CT scans

Image-Guided Radiation Therapy (IGRT) is one of the most effective technologies for the treatment of lung cancer. However, due to respiratory motion, it is crucial to construct a respiratory motion model by registering the images of different phases acquired from 4D-CT to a template image in advance. The treatment plan can be adjusted according to this model and the treatment effect will be improved.

In this experiment, we investigate the registration performance of A-SRWCR on ten thoracic CT scan pairs between extreme inhale and exhale phase of a respiratory cycle, provided by the DIR database (https://www.dir-lab.com/). The scans have a slice thickness of 2.5 mm and an axial resolution ranging from 0.97 to 1.16 mm. Each DIR image includes 300 anatomical landmarks manually annotated by clinical experts. The mean target registration error (mTRE) between these landmarks is severed as the quantitative index. It should be noted that the changes in lung volume due to ventilation are expressed as the differences in corresponding voxel values during the respiratory cycle [29]. This means that the intensities of air and vessels between extreme inhale and exhale phase can suffer from distinct changes [30]. Therefore, we not only need to correct location deformations, but also should take large intensity distortion into account.

Table IV shows the final mTRE results for the ten cases obtained by the four metrics. We also include the results reported in a recent literature [30] which focuses on aligning images with strong intensity distortion. It is observed that A-SRWCR generates the best results for all cases except case #7 and significantly decreases the average mTRE from $8.46 \pm 3.33$ mm to $1.66 \pm 0.53$ mm, which outperforms the second best result ($1.89 \pm 0.89$ mm) by 12%. This improvement shows that A-SRWCR is more reliable in registering images with distinct intensity changes. Especially for case #8, A-SRWCR successfully corrects the deformations, which is far better than others. It verifies the advantages of our method in recovering large deformations. Although all metrics greatly reduce the errors between the landmarks for each case, the three-channel metrics have a more clear decrease compared with MI ($2.60 \pm 1.35$ mm). It confirms the effective performance of this three-channel strategy to intensity fluctuations. Additionally, we also notice that SEMI gives the second highest mTRE value ($2.34 \pm 1.21$ mm). The main reason is that SEMI requires more samples in each subregion to ensure the reliable estimations. Consequently, it is more sensitive to large intensity bias.

A representative example of the aligned slices using different metrics has been visualized in Fig. 4. To display lung vessels clearly, the image intensities are clipped to [50, 750] HU interval. Fig. 4(a) and (b) severally show a slice from the fixed image and the corresponding slice from the moving image, while Fig. 4(c)-(f) illustrate the corresponding slices after registration by the four metrics. The region of interest (ROI) in each slice is marked by a red rectangle and magnified 2 times. Compared with ROI in the fixed slice (Fig. 4(a)), MI (Fig. 4(c)) fails to recover the deformations of the middle vascular branch, while RaPTOR (Fig. 4(e)) provides an excrescent vascular branch in the upper-right corner, pointed to by the yellow dashed circle. Moreover, SEMI (Fig .4(d)) yields a much longer vessel in the upper area. As shown in Fig. 4(f), A-SRWCR obviously improves the registration accuracy of these subtle tissues. It further indicates that our method is more robust to intensity distortion and better in capturing the deformations of small vessel structures.

TABLE IV
QUANTITATIVE ANALYSIS OF 4D-CT DATASET

| Case | TRE [mm] for each method | | | | | |
|---|---|---|---|---|---|---|
| | Initial | MI | SEMI | RaPTOR | Alost[30] | A-SRWCR |
| #1 | 3.89 | 1.23 | 1.06 | 1.09 | – | **1.00** |
| #2 | 4.34 | 1.15 | 1.07 | 0.99 | – | **0.95** |
| #3 | 6.94 | 1.92 | 1.87 | 1.89 | – | **1.34** |
| #4 | 9.83 | 1.81 | 1.78 | 1.54 | – | **1.40** |
| #5 | 7.48 | 2.52 | 2.38 | 2.23 | – | **1.78** |
| #6 | 10.89 | 2.42 | 2.04 | 1.95 | – | **1.66** |
| #7 | 11.03 | 3.92 | 3.19 | **1.96** | – | 2.46 |
| #8 | 14.99 | 5.65 | 5.19 | 3.62 | – | **1.84** |
| #9 | 7.92 | 3.05 | 2.89 | 2.92 | – | **2.50** |
| #10 | 7.30 | 2.29 | 1.96 | 1.89 | – | **1.69** |
| Ave | 8.46 | 2.60 | 2.34 | 2.01 | 1.89 | **1.66** |
| Std | 3.33 | 1.35 | 1.21 | 0.79 | 0.89 | **0.53** |

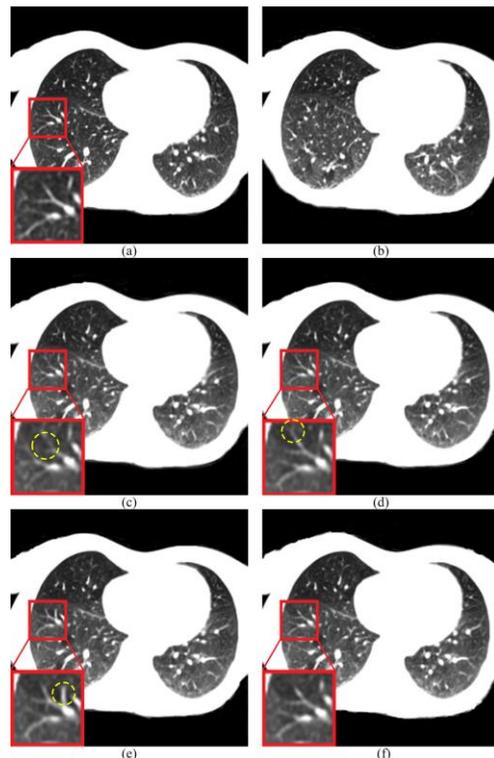

Fig. 4. A slice from the registration pair of case #1. (a) is the slice from the fixed scans; (b) is the corresponding slice from the moving scans; (c), (d), (e) and (f) are the corresponding slices from the aligned scans obtained by MI, SEMI, RaPTOR and A-SRWCR, respectively.

### C. Registration of 3-D OCT frames

Retinal OCT is a non-invasive imaging technology, and the longitudinal registration between multiple OCT images acquired from the same subject at different time allows for monitoring the development and assessing the efficacy of many eye diseases. However, due to the coherent detection characteristics, OCT images are accompanied with strong speckle noise and inherently own a low signal-to-noise ratio (SNR) [31]. Fig. 5 illustrates a B-scan view of 3-D OCT image, in which most voxels are background polluted by strong noise.

TABLE V
QUANTITATIVE ANALYSIS OF 3D OCT DATASET

| Case | MHD [μm] obtained by different methods for each case | | | | | HD [μm] obtained by different methods for each case | | | | |
|---|---|---|---|---|---|---|---|---|---|---|
| | Initial | MI | SEMI | RaPTOR | A-SRWCR | Initial | MI | SEMI | RaPTOR | A-SRWCR |
| #1 | 49.46 | 39.44 | 42.35 | 33.29 | **31.90** | 268.58 | 239.78 | 210.21 | 220.35 | **183.11** |
| #2 | 108.61 | 51.30 | 49.02 | 36.53 | **34.21** | 436.42 | 498.14 | **275.77** | 283.52 | 296.49 |
| #3 | 82.82 | 35.59 | 40.73 | 37.94 | **34.68** | 419.54 | 235.16 | 231.15 | 280.09 | **200.58** |
| #4 | 702.18 | 53.18 | 60.49 | 64.87 | **40.07** | 1796.23 | 387.88 | 731.34 | 700.67 | **262.93** |
| #5 | 56.47 | 37.64 | 39.47 | 38.22 | **33.72** | 264.28 | 292.38 | 238.52 | 319.89 | **163.52** |
| #6 | 114.69 | 55.38 | 75.81 | 51.15 | **39.94** | 625.79 | 401.25 | 512.81 | 578.14 | **301.46** |
| #7 | 125.82 | 60.03 | 72.39 | 48.05 | **45.14** | 584.59 | 510.46 | 592.63 | 415.45 | **387.47** |
| #8 | 171.53 | 46.27 | 44.85 | 54.97 | **33.26** | 520.67 | 368.47 | 341.09 | 649.63 | **335.82** |
| #9 | 68.83 | 35.36 | 35.1 | 35.33 | **29.21** | 335.25 | 338.62 | **204.96** | 362.30 | 264.31 |
| #10 | 105.75 | 31.28 | 33.23 | 29.69 | **27.62** | 427.41 | 201.52 | 187.68 | 275.38 | **165.37** |
| Ave | 158.62 | 44.55 | 49.34 | 43.00 | **34.98** | 567.88 | 347.37 | 352.62 | 408.54 | **256.11** |
| Std | 194.39 | 9.97 | 15.13 | 11.22 | **5.33** | 448.40 | 106.91 | 191.43 | 172.42 | **76.47** |

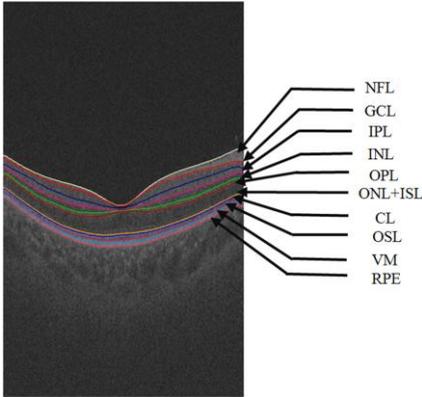

Fig. 5. A slice of 3-D OCT image and 10 retinal layers with 11 surfaces.

In this experiment, we investigate the performance of the four metrics on low SNR retinal OCT dataset, collected from ten healthy volunteers using Topcon 3D-OCT 1000 scanner. Each volunteer collects two 3D OCT sets at interval of one month. The scans are comprised of $512\times 992\times 256$ voxels with a resolution of $11.72\times 2.62\times 23.44$ μm$^3$. Each image is automatically segmented into 10 retinal layers with 11 surfaces (see Fig. 5) by using a graph search based surface detection algorithm [32], and the segmentation accuracy is identified by ophthalmologists. For each pair (the same volunteer collected at different time), the mean HD and MHD values of 11 surfaces after registration are employed to quantify the alignment accuracy. Due to the limitation of GPU memory, we first register two down-sampling images with dimension $512\times 512\times 128$, and then up-sample the deformation fields to the original resolution.

As listed in Table V, A-SRWCR produces the lowest MHD for all cases, and greatly reduces the average MHD from $158.62\pm 194.39$ μm to $34.98\pm 5.33$ μm. Although the HD values in cases #2 and #9 are slightly higher than the optimal results, A-SRWCR yields a statistically significant improvement with the lowest average HD ($256.11\pm 76.47$ μm). It demonstrates that A-SRWCR is more robust to speckle noise. Especially for case #4, in which the initial OCT images are completely mismatched, the outcomes obtained by A-SRWCR are still around the average values, while all contrast algorithms converge to the invalid local optima. It further confirms that A-SRWCR is more potential to correct large deformations even in the presence of strong speckle noise. There are two reasons for the excellent performance of our proposed metric. First, CR estimates an optimal function to map voxel pairs with the smallest distance in intensity space, which is verified to be less sensitive to noise [33]. Second, compared with the boxcar function, the cubic B-spline function preserves the topological relationship between the voxels within each subregion. It is therefore more reliable to find the optimal functional relationship even if the region is seriously polluted by noise.

Additionally, it might be confusing that the performance of SEMI is even worse than MI. The reason is that most subregions contain little retinal structural information, and some may even be all noisy. Consequently, the local MI values estimated over these subregions do not reflect the intensity bin correspondence of retinal tissues.

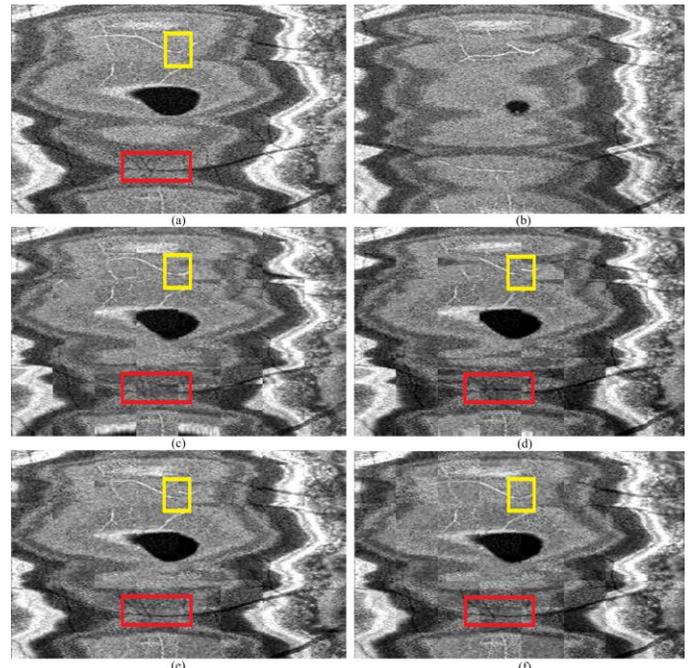

Fig. 6. Checkerboard images with a C-scan view using the four metrics. (a) is the slice from the fixed scans; (b) is the corresponding slice from the moving scans; (c), (d), (e) and (f) are the corresponding checkerboard images obtained by MI, SEMI, RaPTOR and A-SRWCR, respectively.

TABLE VI
QUANTITATIVE ANALYSIS OF CT/PET DATASET

| Case | MHD [mm] obtained by different methods for each case | | | | | HD [mm] obtained by different methods for each case | | | | |
|---|---|---|---|---|---|---|---|---|---|---|
| | Initial | MI | SEMI | RaPTOR | A-SRWCR | Initial | MI | SEMI | RaPTOR | A-SRWCR |
| #1 | 10.30 | 6.71 | 6.07 | 4.72 | **4.22** | 62.54 | 60.55 | 65.90 | 46.45 | **28.91** |
| #2 | 13.25 | 5.12 | 4.27 | 5.30 | **3.71** | 70.13 | 36.23 | 36.19 | 29.69 | **28.82** |
| #3 | 4.12 | 3.32 | 3.28 | **2.59** | 3.14 | 30.77 | 31.26 | 34.42 | 27.47 | **25.42** |
| #4 | 7.52 | **3.67** | 4.14 | 4.33 | 4.12 | 43.91 | 30.58 | 31.97 | 26.83 | **23.58** |
| #5 | 11.91 | 5.34 | 4.49 | 4.93 | **4.47** | 71.38 | 45.42 | 48.82 | 44.89 | **40.95** |
| #6 | 8.79 | 4.96 | 4.52 | 4.04 | **4.02** | 78.01 | 44.09 | 49.68 | **30.03** | 31.85 |
| #7 | 5.66 | **2.93** | 3.43 | 3.02 | 3.37 | 36.45 | 25.82 | 26.56 | **25.66** | 27.34 |
| #8 | 10.03 | 3.80 | 3.24 | 3.40 | **2.75** | 58.19 | **26.53** | 27.07 | 27.44 | 27.38 |
| #9 | 7.72 | 3.66 | 3.24 | **3.16** | 3.56 | 47.94 | 32.16 | 39.49 | 32.49 | 36.36 |
| #10 | 9.26 | 5.70 | 5.64 | 4.41 | **3.63** | 54.07 | 54.93 | 63.31 | 43.77 | **31.87** |
| Ave | 8.86 | 4.52 | 4.23 | 3.99 | **3.70** | 55.34 | 38.76 | 42.34 | 33.47 | **30.25** |
| Std | 2.75 | 1.22 | 1.00 | 0.90 | **0.52** | 15.63 | 12.42 | 14.10 | 8.94 | **5.22** |

To make the results of A-SRWCR more convincible, a qualitative comparison of the C-scan view which contains retinal vessels is illustrated in Fig. 6. Fig. 6(a) and (b) are the fixed and corresponding moving slices, respectively. Fig. 6(c)-(f) show the checkerboard images which alternately arrange the fixed and registered slices. It can be seen that SEMI (Fig. 6(d)) gives the worst registration result since the vessels in both two ROIs are broken. Moreover, MI (Fig. 6(c)) provides a good alignment of the vessel within the yellow ROI, but completely fails in the red ROI. In contrast, RaPTOR (Fig. 6(e)) provides a more continuous vessel within the red ROI, but fails in the yellow ROI. As shown in Fig. 6(f), A-SRWCR clearly improves the registration accuracy of the retinal vessels, thus further verifying that A-SRWCR is more suitable to register images with strong speckle noise.

### D. Registration of lung CT/PET images

Non-rigid multi-modal registration is also an insurmountable obstacle for many applications. In this experiment, we apply our proposed method to a clinical lung database collected from ten patients. Each patient has been scanned by PET-CT and high quality diagnostic CT at different time. Due to the different scanning protocols, the diagnostic CT scans acquired from Siemens Somatom have a slice thickness of 5 mm and an in-plane resolution around 1.20-1.46 mm, while the PET scans acquired from the integrated PET-CT scanner Siemens Biograph 64 have a resolution of $4.07 \times 4.07 \times 3$ mm$^3$. To improve the registration efficiency, each pair has been resampled to a uniform resolution with 3 mm slices and the same in-plane resolution as CT image. Moreover, a rigid registration of all cases using A-SRWCR with the rigid transformation model is performed to recover global deformations. Since it is quite difficult to manually select landmarks that can be observed both in CT and PET images, the MHD and HD values between the lung surfaces are provided for quantitative comparison of the registration accuracy.

Table VI shows the quantitative results achieved by the four metrics. Although the MHD or HD values obtained by A-SRWCR are not the lowest for some cases, such as case #7 or #9, our method has a statistically significant improvement over the ten cases, and achieves the lowest average and standard deviation values ($3.70 \pm 0.52$ mm for MHD, $30.25 \pm 5.22$ mm for HD). It demonstrates that A-SRWCR is more stable to align CT/PET images. Especially for cases #1 and #10, the comparative methods completely fail to align the point pairs with maximal displacement, whereas A-SRWCR greatly reduces the HD values. Additionally, we also find that the registration performance of the MI-based metrics is much worse than the CR-based metrics. In CT images, there is one intensity bin for several structures, but the intensities of these tissues might correspond to a number of intensity bins in PET images. Consequently, the optimal alignment of these tissues do not obey the hypothesis of the intensity bin correspondence. In contrast, RaPTOR and A-SRWCR quantify the functional dependence between the intensity values instead of the intensity bins. They are therefore more potential to register CT/PET images. However, due to the approximate reductions, RaPTOR is less stable and yields larger standard deviation values.

Fig. 7 illustrates a representative example of fusion images with a coronal view, in which the PET images have been enhanced with green for better observation. By comparing the ROI marked by the red rectangle, it is obvious that A-SRWCR provides more accurate alignment of the lung diaphragm than the other three methods.

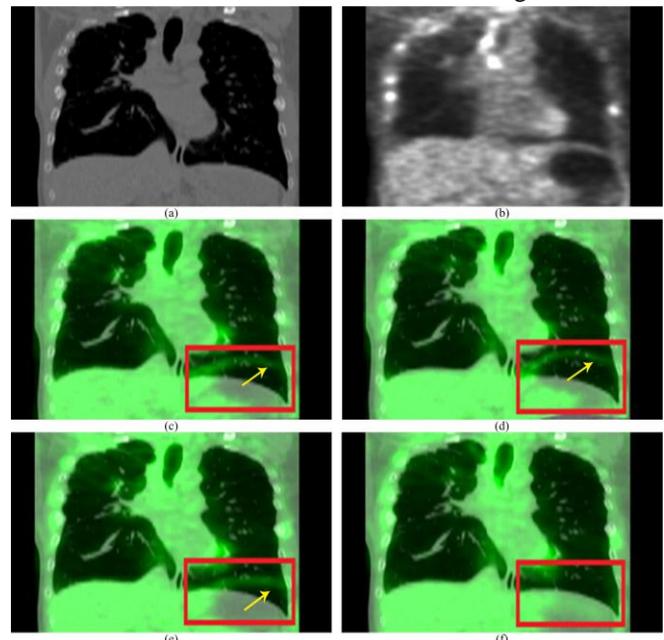

Fig. 7. Visualization of the fusion images with a coronal view. (a) is the slice from CT scans; (b) is the corresponding slice from PET scans; (c), (d), (e) and (f) are the corresponding fusion images obtained by MI, SEMI, RaPTOR and A-SRWCR, respectively.

*E. Experiments of GPU Acceleration*

In this experiment, we validate the acceleration performance of the parallel mechanism. According to our proposed framework, there are four CUDA kernels performed on the GPU during each iteration. First, kernel 1 reads the initial transformation parameters and the moving image from the RAM, and outputs the transformed image to the RAM. Then, kernel 2 reads the fixed image from the RAM and outputs the regional joint PDFs to the RAM. Third, kernel 3 accomplishes the computation of $\partial D/\partial M(\mathbf{y})$ and $\partial M(\mathbf{y})/\partial \mathbf{y}$ for each voxel. Finally, kernel 4 outputs the gradient of SRWCR to the RAM.

Table VII shows the memory bandwidth occupied by each kernel at the finest deformation level. The results are measured by Visual Studio Nsight. For each kernel, we reach around half of the theoretical peak bandwidth, which is 192.2 GB/s.

TABLE VII
QUANTITATIVE ANALYSIS OF MEMORY BANDWIDTH

| kernel | 1 | 2 | 3 | 4 |
|---|---|---|---|---|
| Bandwidth (GB/s) | 94.31 | 81.46 | 108.99 | 95.28 |
| Efficiency (%) | 49.07 | 42.38 | 56.71 | 49.57 |

Especially for kernel 3, the efficiency is up to 56.7%.

Table VIII lists the computational time of SRWCR and its derivatives taken by the CPU and GPU during one iteration with different image sizes. The calculation speed of our paralleled scheme has a significant improvement compared with that of the CPU. It might be puzzling that the acceleration rate of SRWCR is much lower than that of its derivatives. The main reason is due to the atomic operations, which enforce the threads serialize access to the same address and seriously diminish the parallel efficiency.

Fig. 8 illustrates the average registration time for four datasets. Considering that the public 4D-CT dataset contains two different in-plane dimensions ($256\times256$ and $512\times512$), we compute each average value separately. Relying on the dataset to be registered, execution time consumed by our method ranges from 3 to 37.2 minutes, which is approximately 24 times less than the CPU-based registration. It successfully confirms the impact of our parallel mechanism.

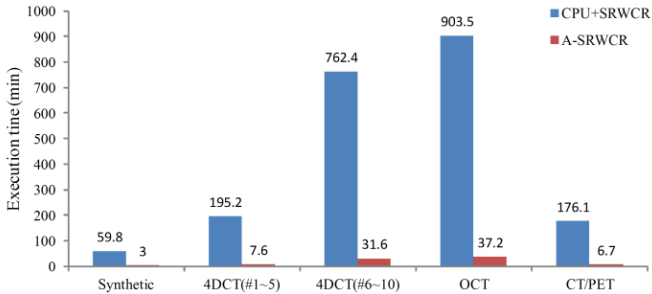

Fig. 8. Average execution time for four datasets.

## IV. DISCUSSION

MI is an extensively used similarity measure which aims to maximize the amount of shared information between two images. However, when confronted with practical challenges such as intensity distortion or different imaging modalities, it may lead to undesired results. SEMI and RaPTOR are two state-of-the-art techniques which encode the spatial information into the statistical intensity relationship and greatly improve the robustness to these challenges. But there are still some drawbacks for both two methods. SEMI is sensitive to the size of subregion, and thereby its local performance is significantly limited. RaPTOR, which assumes that the spatial distribution is corresponding to a boxcar function, can not differentiate the contribution of voxels within each subregion.

In this work, we introduce A-SRWCR, a novel similarity measure that incorporates the spatial information into the functional mapping relationship. A-SRWCR starts from a 3-D PDF which extends the intensity dimensions with a spatial channel, and then estimates local CR values give the spatial distribution modeled by a cubic B-spline function. We also design an efficient parallel mechanism to overcome the shortcoming of huge computation burden.

We perform an extensive comparison of our proposed method and the three metrics: MI, SEMI and RaPTOR. Experiments with synthetic images provide the evaluation of both accuracy and robustness to bias fields. Application in DIR dataset demonstrates the effectiveness of A-SRWCR in registering images with distinct intensity distortion. The optimal alignment of lung vessels shown in Fig. 4 further demonstrates that A-SRWCR could provide a more accurate respiratory motion model for IGRT. Experiments with OCT scans show an objective comparison of the robustness of the four metrics to speckle noise. The lowest HD and MHD values listed in Table V imply that A-SRWCR is more reliable in analyzing the thickness variations of retinal layers, which is specifically useful in the surveillance of multiple sclerosis (MS) [34]. The highest accuracy illustrated in Fig. 6 also verifies that the presented metric is more suitable for the assessment of choroidal neovascularization (CNV), an eye disease caused by vasculopathy. For multi-modal registration validated with clinical CT/PET scans, A-SRWCR also provides a statistically significant improvement over all other three metrics.

Several reasons account for the excellent performance of A-SRWCR. First, similar to RaPTOR, A-SRWCR is a local similarity measure. Although it may be over-constrained to map all intensities between two images using one function, it is more feasible to assume a functional dependence between intensities within a subregion, especially for small region. Second, in virtue of the cubic B-spline function, the intensity of each voxel is weighted in terms of the distance to the center of subregion. It is therefore more reliable in estimating one-to-one mapping relationship for each region. Third, we rigorously deduce A-SRWCR from a three-channel joint PDF which is estimated by a second-order polynomial function. It ensures that the presented metric is more accurate to measure the difference between two images and more differentiable.

TABLE VIII
COMPUTATION TIME DURING AN ITERATION FOR DIFFERENT IMAGE SIZE.

| Image size | SRWCR | | | Derivatives | | |
|---|---|---|---|---|---|---|
| | CPU (ms) | GPU (ms) | Acceleration rate | CPU (ms) | GPU (ms) | Acceleration rate |
| 64×64×24 | 327 | 21 | 15.6 | 297 | 4 | 74.3 |
| 128×128×49 | 2278 | 132 | 17.3 | 2153 | 32 | 67.3 |
| 256×256×99 | 18533 | 851 | 21.8 | 16989 | 238 | 71.4 |

Recently, the symmetric diffeomorphic transformation model which ensures the reversible spatial deformations has been successfully incorporated into some similarity measures and significantly improves the registration accuracy [35], [36]. In the future, we will investigate the potential advantages of combining A-SRWCR with this symmetric diffeomorphic framework on more challenging applications such as the diseased OCT images.

## V. CONCLUSION

We have presented SRWCR as a novel similarity measure for non-rigid registration, and sped up the computation of SRWCR and its derivatives using CUDA-programming. The experiments on both mono-modal and multi-modal datasets have shown that A-SRWCR is very stable and outperforms the existing methods such as MI, SEMI or RaPTOR. With more accurate matching performance and higher speed, our proposed registration framework is more suitable to handle the clinical challenges.

## APPENDIX I

Here, we derive (19) from (18) in detail. The key is to compute the derivatives of $\sigma_r^2$, $\mu_r(a)$ with the respect to $B(\mathbf{y})$. Therefore, we first compute them independently.

$$\frac{\partial \sigma_r^2}{\partial B(\mathbf{y})} = \sum_{b=0}^{L_\varepsilon} b^2 \frac{\partial p_r(b)}{\partial B(\mathbf{y})} - 2\mu_r \frac{\partial \mu_r}{\partial B(\mathbf{y})}$$

$$= \sum_{b=0}^{L_\varepsilon} b^2 \frac{\partial p_r(b)}{\partial B(\mathbf{y})} - 2\mu_r \sum_{b=0}^{L_\varepsilon} b \frac{\partial p_r(b)}{\partial B(\mathbf{y})}$$

$$= \sum_{b=0}^{L_\varepsilon} (b^2 - 2\mu_r b) \frac{\partial \sum_{a=0}^{L_\varepsilon} p_r(a,b)}{\partial B(\mathbf{y})} . \quad (22)$$

Since $p_r(a)$ is independent in $B(\mathbf{y})$, $\partial \mu_r(a)/\partial B(\mathbf{y})$ can be calculated as follows

$$\frac{\partial \mu_r(a)}{\partial B(\mathbf{y})} = \frac{1}{p_r(a)} \sum_{b=0}^{L_\varepsilon} b \frac{\partial p_r(a,b)}{\partial B(\mathbf{y})} . \quad (23)$$

Combining (22) and (23) with (18), we obtain

$$\frac{\partial D}{\partial M(\mathbf{y})} = \sum_{r=0}^{R_n} p(r) \left[ \frac{1}{\sigma_r^2} \left( \sum_{b=0}^{L_\varepsilon} b^2 \frac{\partial \sum_{a=0}^{L_\varepsilon} p_r(a,b)}{\partial B(\mathbf{y})} - \sum_{a=0}^{L_\varepsilon} 2\mu_r(a) \sum_{b=0}^{L_\varepsilon} b \frac{\partial p_r(a,b)}{\partial B(\mathbf{y})} \right) \right.$$

$$\left. - \frac{(1 - CR(A,B|r))}{\sigma_r^2} \sum_{b=0}^{L_\varepsilon} (b^2 - 2b\mu_r) \frac{\partial \sum_{a=0}^{L_\varepsilon} p_r(a,b)}{\partial B(\mathbf{y})} \right]. \quad (24)$$

After merging the same items, (24) can be simplified as

$$\frac{\partial D}{\partial M(\mathbf{y})} = \sum_{r=0}^{R_n} \sum_{a=0}^{L_\varepsilon} \sum_{b=0}^{L_\varepsilon} \left[ \frac{b^2 - 2b\mu_r(a) - (1 - CR(A,B|r))(b^2 - 2b\mu_r)}{\sigma_r^2} \frac{\partial p(a,b,r)}{\partial B(\mathbf{y})} \right]. \quad (25)$$

In virtue of the chain rule, the derivative of the 3-D joint PDF defined by (3) can be deduced as follows

$$\frac{\partial p(a,b,r)}{\partial B(\mathbf{y})} = -\frac{1}{Z} w(r,\mathbf{x}) h(a - A(\mathbf{x})) \frac{\partial h}{\partial \kappa}\big|_{\kappa = b - B(\mathbf{y})} . \quad (26)$$

Combining (25) with (26), we have

$$\frac{\partial D}{\partial M(\mathbf{y})} = \frac{1}{Z} \sum_{r=0}^{R_n} \sum_{a=0}^{L_\varepsilon} \sum_{b=0}^{L_\varepsilon} \left[ \frac{(1 - CR(A,B|r))(b^2 - 2b\mu_r) + 2b\mu_r(a) - b^2}{\sigma_r^2} \right.$$

$$\left. \cdot w(r,\mathbf{x}) h(a - A(\mathbf{x})) \frac{\partial h}{\partial \kappa}\big|_{\kappa = b - B(\mathbf{y})} \right]. \quad (27)$$

## APPENDIX II

According to (20), we first compute $\partial \mu_r(a)/\partial A(\mathbf{y})$ as follows

$$\frac{\partial \mu_r(a)}{\partial A(\mathbf{y})} = -\frac{\partial p_r(a)/\partial A(\mathbf{y})}{p_r^2(a)} \sum_{b=0}^{L_\varepsilon} b p_r(a,b) + \frac{1}{p_r(a)} \sum_{b=0}^{L_\varepsilon} b \frac{\partial p_r(a,b)}{\partial A(\mathbf{y})}$$

$$= -\frac{\partial \sum_{b=0}^{L_\varepsilon} p_r(a,b)/\partial A(\mathbf{y})}{p_r(a)} \mu_r(a) + \frac{1}{p_r(a)} \sum_{b=0}^{L_\varepsilon} b \frac{\partial p_r(a,b)}{\partial A(\mathbf{y})} . \quad (28)$$

Incorporating (28) into (20), $\partial D/\partial M(\mathbf{y})$ can be re-written as

$$\frac{\partial D}{\partial M(\mathbf{y})} = \sum_{r=0}^{R_n} \frac{-p(r)}{\sigma_r^2} \left[ \sum_{a=0}^{L_\varepsilon} \left( -2\mu_r^2(a) \frac{\partial \sum_{b=0}^{L_\varepsilon} p_r(a,b)}{\partial A(\mathbf{y})} + \mu_r(a) \sum_{b=0}^{L_\varepsilon} b \frac{\partial p_r(a,b)}{\partial A(\mathbf{y})} + \mu_r^2(a) \frac{\partial \sum_{b=0}^{L_\varepsilon} p_r(a,b)}{\partial A(\mathbf{y})} \right) \right]. \quad (29)$$

Similar to (25), (29) can be simplified as follows

$$\frac{\partial D}{\partial M(\mathbf{y})} = \sum_{r=0}^{R_n} \sum_{a=0}^{L_\varepsilon} \sum_{b=0}^{L_\varepsilon} \left[ \frac{\mu_r^2(a) - 2b\mu_r(a)}{\sigma_r^2} \frac{\partial p(a,b,r)}{\partial A(\mathbf{y})} \right] . \quad (30)$$

Combining (30) with (25), we obtain

$$\frac{\partial D}{\partial M(\mathbf{y})} = \frac{1}{Z} \sum_{r=0}^{R_n} \sum_{a=0}^{L_\varepsilon} \sum_{b=0}^{L_\varepsilon} \left[ \frac{2b - \mu_r(a)}{\sigma_r^2} \mu_r(a) w(r,\mathbf{x}) h(b - B(\mathbf{x})) \frac{\partial h}{\partial \kappa}\big|_{\kappa = a - A(\mathbf{y})} \right]. \quad (31)$$

## ACKNOWLEDGMENT

The authors would like to thank MOSTC, NSFC, NSFJ and CAS for financial support, colleague Chengtao Peng for valuable discussions and Yang Yang from Huashan Hospital for providing the clinical CT and PET images. The authors are also grateful to anonymous reviewers for their valuable comments and suggestions. The authors declare that there are no financial interests of the authors in any company.